\documentclass[prd,aps,secnumarabic,superscriptaddress]{revtex4}
\usepackage{graphicx}
\usepackage{bm}
\usepackage{amsmath}
\usepackage{amsfonts}
\usepackage{amssymb}
\def\Dot{\!\cdot\!}

\def\ds{\displaystyle}

\def\sb{\mbox{\rule[-10pt]{0pt}{26pt}}}
\def\sc{\mbox{\rule[-4pt]{0pt}{14pt}}}
\def\ep{\varepsilon}

\def\be{\begin{equation}}
\def\ee{\end{equation}}
\def\bea{\begin{eqnarray}}
\def\eea{\end{eqnarray}}
\def\bc{\begin{displaymath}}
\def\ec{\end{displaymath}}
\def\la{\lambda}
\def\Li2{\mathrm{Li_2}}
\begin{document}
\title{Noncommutative QED corrections to ${\bm {e^+e^-\to\gamma\gamma\gamma}}$ at linear collider energies}
\author{Alberto Devoto}
\email{Alberto.Devoto@ca.infn.it}
\affiliation{Dipartimento di Fisica, Universit\`a di Cagliari and INFN, Sezione di Cagliari, Cagliari, Italy} 
\author{Stefano Di Chiara}
\email{dichiara@pa.msu.edu}
\affiliation{Dipartimento di Fisica, Universit\`a di Cagliari and INFN, Sezione di Cagliari, Cagliari, Italy}
\affiliation{Department of Physics and Astronomy, Michigan State University, East Lansing, MI 48824}
\author{Wayne W. Repko}
\email{repko@pa.msu.edu} 
\affiliation{Department of Physics and Astronomy, Michigan State University, East Lansing, MI 48824}

\date{\today}

\begin{abstract}
We compute the total cross section as well as angular and energy distributions for process $e^+e^-\to\gamma\gamma\gamma$ with both unpolarized and polarized beams in the framework of noncommutative quantum electrodynamics (NCQED). The calculation is performed in the center of mass of colliding electron and positron and is evaluated for energies and integrated luminosities appropriate to future linear colliders. We find that by using unpolarized beams it is possible to probe the Lorentz symmetry violating azimuthal dependence of the cross section. Furthermore, with polarized beams the left-right asymmetry of the CP violating NCQED amplitudes can be used to obtain bounds on the noncommutative scale $\Lambda_{NC}$ which exceed 1.0 TeV.
\end{abstract}
\maketitle

\section{Introduction}

The idea of formulating field theories on noncommutative spaces goes back some time \cite{snyder}. Interest has been revived recently with the realization that noncommutative quantum field theories emerge in the low energy limit of string theories \cite{cds,dh,sw}. This has led to numerous investigations of the phenomenological implications of noncommutative QED \cite{HPR,pheno}.

In noncommutative geometries, the coordinates $x^{\mu}$ obey the commutation relations
\begin{equation}
      \left[ x^{\mu} , x^{ \nu} \right] = i \theta^{\mu\nu},
\end{equation}
where $\theta^{\mu\nu}=-\theta^{\nu\mu}$. 
The extension of quantum field theories from ordinary space-time to noncommutative space-time is achieved replacing the ordinary products with Moyal $\star$ products, defined by
\begin{equation} \label{Moyal}
\left(f \star g \right)\left(x \right)= \exp \left({\scriptstyle\frac{1}{2}}i\theta^{\mu \nu} \partial_{x^{\mu}} \partial_{y^{\nu}} \right)f \left(x \right)g \left(y \right)\big| _{x=y} {\ .}
\end{equation}
Here, in order to ensure the S matrix unitarity, we assume that $\theta^{0 i}=\theta^{i 0}=0$,  
$i=1,2,3$, .

In the following, we study the effect of noncommutative geometry on the process 
$e^{+}\!e^{-}\ \rightarrow\ \gamma\gamma\gamma$ using noncommutative quantum electrodynamics, NCQED. NCQED, defined and described for example in \cite{NcQED}, has as its Lagrangian 
\begin{equation} \label{Lag}
      \mathcal{L}= \bar{\psi}\star \left(\not\!\! D- m \right)\psi - \frac{1}{2e^{2}}%
 \mathrm{Tr}\left(F_{\mu \nu} \star F^{\mu\nu} \right)+\mathcal{L}_{gauge}+\mathcal{L}_{ghost}\,,
\end{equation}
where $\mathcal{L}_{gauge}$ and $\mathcal{L}_{ghost}$ denote the gauge fixing and ghost terms. The corresponding Feynman rules for phenomenological calculations can be derived from Eq.\,(\ref{Lag}) \cite{NcQED}. Since scale at which noncommutative effects are likely to occur is large, we focus on the energy scales typically associated with future linear colliders. Our calculations are performed in the center of mass of the colliding electron and positron. In the next section, we outline the calculation of the squared amplitudes. This is followed by a summary of the cross section computations and a discussion of the results. Details of the calculation are presented in the Appendices.

\section{NCQED Amplitudes}

\begin{figure}[h]
\hfill\includegraphics[height=1.4in]{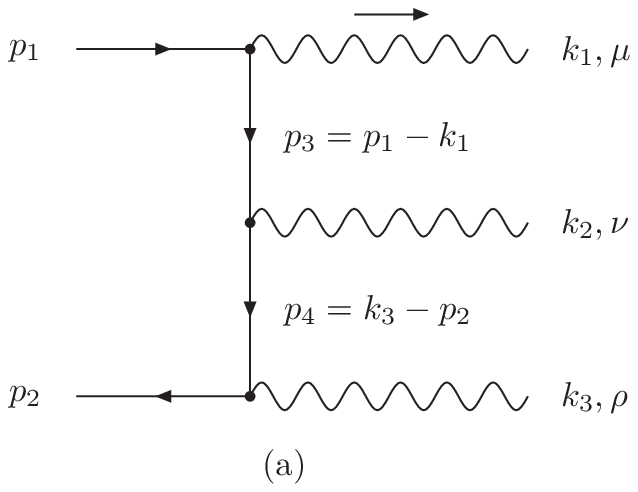}
\hfill\includegraphics[height=1.4in]{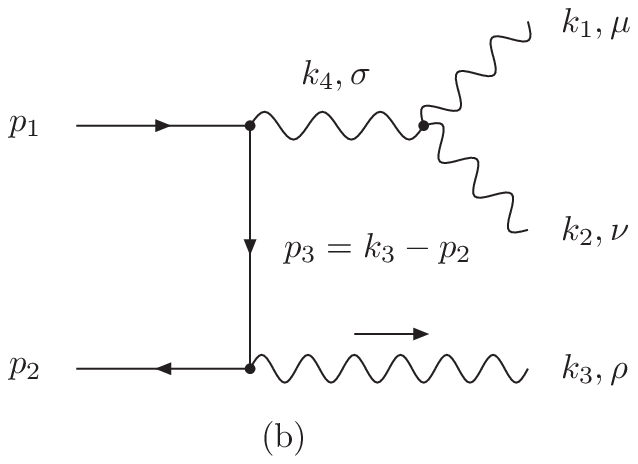}\hfill\includegraphics[height=1.4in]
{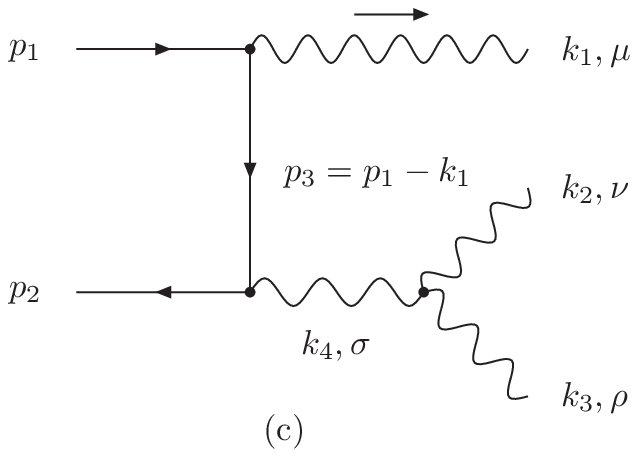}\hspace*{0pt\hfill}
\caption{Diagram (a) represents the contributions from abelian interaction terms and diagrams (b) and (c) represent contributions from the non-abelian interaction terms. \label{diags}}
\end{figure}

Typical Feynman graphs contributing to the $e^+e^-\to\gamma\gamma\gamma$ at leading order are shown in Fig.\,\ref{diags}. Diagrams in which the fermion line is connected to the final photons by a single photon propagator vanish. The complete set is obtained by permuting the photons, which gives a total of twelve diagrams. We have calculated specific helicity amplitudes and it is therefore unnecessary to include ghost contributions \cite{GW}.  
The amplitudes were calculated using the Feynman rules in Ref.\cite{NcQED}. We computed the helicity amplitudes with the aid of the symbolic manipulation program FORM and simplified the results using \emph{Mathematica}.

To simplify the presentation of the results, we introduce kinematical variables  
\begin{eqnarray}
      a=k_2\Dot k_3 b&=&k_3\Dot k_1 c=k_1\Dot k_2 \label{kine1}\\
      p=p_1\Dot k_1 q&=&p_1\Dot k_2 r=p_1\Dot k_3 \label{kine2}\\
      s=p_2\Dot k_1 t&=&p_2\Dot k_2 u=p_2\Dot k_3 \label{kine3}\\
v=\cos\left(k_1\Dot \theta\Dot k_2/2\right)&\hspace{10pt}& w=\sin\left(k_1\Dot \theta\Dot k_2/2\right) \label{kine4} \\
      S&=&\left(p_1+p_2\right)^2\,,
\end{eqnarray}
where $p\Dot\theta\Dot q=p_\mu\theta^{\mu\nu}q_\nu$.
These variables are related as
\begin{equation}
\begin{array}{rcccl}
      S/2&=&s+t+u&=&p+q+r \\
      a&=&q+r-s&=&t+u-p \\
      b&=&p+r-t&=&s+u-q \\
      c&=&p+q-u&=&s+t-r\ .
\end{array}
\end{equation}
In the center of mass system we also have
\begin{equation} \label{kthk}
k_1\Dot \theta\Dot k_2=k_2\Dot \theta\Dot k_3=k_3\Dot \theta\Dot k_1\ .
\end{equation}
While there are only five independent variables, it is convenient for displaying the results and examining symmetries to retain six. If the helicities are labelled $\lambda,\bar\lambda;\lambda_1,\lambda_2,\lambda_3$, the square of helicity amplitude $+,-;+,+,-$ is
\begin{eqnarray} \label{eliMq}
\left|\mathcal{M}_{ +,-;+,+,- }\right|^2 &=& \frac{2\,e^6\,r^2}{p\,q\,s\,t} \bigg[ S - 4\,w^2\, \bigg(\frac{3\,S}{2}
 + \frac{a\,\left(p^2 + s^2 \right)}{b\,c} + 
     \frac{b\,\left(q^2 + t^2 \right)}{a\,c} + 
     \frac{c\,\left(r^2 + u^2 \right)}{a\,b} \nonumber \\ 
 &-& \frac{p\,q + \left(p + s \right)\,\left(q + t \right)+ s\,t }{c}   
 - \frac{p\,r + \left(p + s \right)\,\left(r + u \right)+ s\,u }{b} 
 - \frac{q\,r + \left(q + t \right)\,\left(r + u \right)+ t\,u }{a}\bigg)\nonumber \\  
 &-& 4\,v\,w\, \left(\frac{1}{a} + \frac{1}{b} + \frac{1}{c} \right)
  \epsilon(k_1,k_2,p_1,p_2)\bigg]\ .
\end{eqnarray}
The remaining squared amplitudes are given in Appendix \ref{Amps}.

With the aid of Eqs.\,(\ref{kine1},\ref{kine2},\ref{kine3},\ref{kine4},\,and\,\ref{kthk}), it is easy to check that Eq.\,(\ref{eliMq}) satisfies Bose symmetry. The last term in Eq.\,(\ref{eliMq}) changes sign under the exchange of $p_1$ and $p_2$, which is a reflection of the lack of charge conjugation symmetry. The parity transformation $\vec{p}\to-\vec{p}$, $\lambda\to-\lambda$ gives $\mathcal{M}_{\lambda,\bar\lambda;\lambda_1,\lambda_2,\lambda_3 } = \mathcal{M}_ {-\lambda,-\bar\lambda;-\lambda_1,-\lambda_2,-\lambda_3}$, as can be seen using Appendix\,\ref{Amps}. As a consequence, CP is violated, but CPT is preserved \cite{CPT}. The CP violating terms cancel in the sum over fermion helicities.

Summing over photon helicities and averaging over electron helicities gives  
\begin{eqnarray} \label{Mq}
\left|\mathcal{M}_{e^+ e^- \rightarrow \gamma\gamma\gamma}\right|^2 &=& e^6\,\frac{ p\,s\,\left(p^2 + s^2 \right)+ q\,t\,\left(q^2 + t^2 \right)+ 
      r\,u\,\left(r^2 + u^2 \right)}{p\,q\,r\,s\,t\,u} \nonumber \\ 
 &\times&\bigg[ S- 4\, w^2\, \bigg(\frac{a\,\left(p^2 + s^2 \right)}{b\,c} + 
                     \frac{b\,\left(q^2 + t^2 \right)}{a\,c} + 
                     \frac{c\,\left(r^2 + u^2 \right)}{a\,b}  
  - \frac{p\,q + \left(p + s \right)\,\left(q + t \right)+ s\,t }{c}\nonumber \\   
 &-& \frac{p\,r + \left(p + s \right)\,\left(r + u \right)+ s\,u }{b} 
  - \frac{q\,r + \left(q + t \right)\,\left(r + u \right)+ t\,u }{a} 
  +\frac{3\,S}{2}\bigg)\bigg] .
\end{eqnarray}
We have checked that this result satisfies the Ward identity $\mathcal{M}_{\mu}k_i^{\mu}=0$ for each photon and that it reduces to the standard QED result \cite{CGTW} if $\theta^{\mu\nu} = 0$.
\phantom{space\\}
\section{Cross Section Results} \label{sec:CrossSection}
\subsection{Unpolarized Cross Section}

The details of obtaining the cross section from $|\mathcal{M}_{e^+ e^- \rightarrow \gamma\gamma\gamma}|^2$, Eq.\,(\ref{Mq}), are given in Appendix \ref{avgXsec}.  The result consists of the pure QED cross section, which contains an infrared divergence that must be regularized, and an infrared finite NCQED correction. To check the validity of our helicity amplitudes, we recalculated the QED spin averaged total cross section by retaining all terms linear in the electron mass squared, $m^2$, in the numerators of Eq.\,(\ref{Mq}). Our result then agrees with that of Berends and Kleiss \cite{Berends}, namely 
\begin{equation} \label{sigQED}
      \sigma^{QED}_{e^+ e^- \rightarrow \gamma\gamma\gamma}=
      \frac{2\alpha^3}{S} \left(\left(\log\rho-1\right)^2\left(\log\omega-1\right)+3\right)\,,
\end{equation}
with
\begin{equation}
      \rho=\frac{S}{m^2}\ \ \ \omega=\frac{E^2}{E_{\rm min}^{\,2}}\,.
\end{equation}

By assuming the non-commutative factor $w^2$ in Eq.\,(\ref{Mq}) is small, we have $w^2=\sin^2\!\!\varphi \cong\varphi^2$, where $\varphi$ is
\begin{equation}
\varphi\equiv{\textstyle\frac{1}{2}}k_1\Dot\theta\Dot k_2
  ={\textstyle\frac{1}{2}}\left(\vec{k_1}\times\vec{k_2}\right) \Dot\vec\theta \,.
\end{equation}
It is then possible to express the non-commutative effects in terms of $\la$, the angle between $\vec{p}_1$ and $\vec{\theta}$ and $z=S/\Lambda_{NC}^2$. The NCQED contribution to the spin averaged cross section has the form
\begin{equation} \label{sigNC}
\sigma^{NCQED}_{e^+ e^- \rightarrow \gamma\gamma\gamma}=\frac{z^2 \alpha^3}{S}\left[\frac{2231}{720}-\frac{\pi^2}{120}-\frac{5}{2}\zeta(3) +\sin^2\la\left(\frac{\pi^2}{80}+\frac{7}{2}\zeta(3)-\frac{148957}{34560}-\frac{7 \log\!\rho}{960}\right)\right]\,.
\end{equation}
If we average over $\la$, the expression for the total cross section becomes
\begin{equation}\label{totXsec}
\sigma_{e^+ e^- \rightarrow \gamma\gamma\gamma}=\frac{ \alpha^3}{S}\bigg[2\left(\log\rho-1\right)^2\left(\log\omega-1\right)+6 +z^2\left(\frac{65219}{69120}-\frac{\pi^2}{480}-\frac{3}{4}\zeta(3)-\frac{7 \log\!\rho}{1920}\right)\bigg]\,.
\end{equation}
With or without the average over $\la$, the effect of non-commutativity on spin averaged total cross section is relatively small since it depends on $z^2$ and has a $\log\rho$ rather than a $\log^2\rho$ dependence. 

To determine if NCQED can be tested using the spin averaged three photon process, we examined the dependence of the cross section on the azimuthal angle $\phi$ of one of the photons. This is a pure NCQED effect since the QED cross section has no such $\phi$-dependence. Before doing this, it is important to determine the range of validity of the approximation $\sin^2\varphi\approx\varphi^2$. From Appendix \ref{avgXsec}, $\varphi^2$ has the form
\begin{equation}
\varphi^2=\frac{z^2}{64}\nu_1^2\nu_2^2\sin^2\alpha\sin^2\beta\,,
\end{equation}
where $\nu_i=E_i/E$ and $\alpha$ and $\beta$ vary between $0$ and $\pi$. Thus, the approximation is always good if $z<8$. Since all the other factors vary between $0$ and $1$, and we integrate over some or all of them in calculating the distributions, a limit on $z$ based on the the average value $\left\langle \varphi^2\right\rangle$ can be useful. This limit is
\begin{equation}
z<\frac{8}{\sqrt{\left\langle \nu_1^2\nu_2^2\sin^2\alpha\sin^2\beta\right\rangle}}<35\,.
\end{equation} 
In the following, we choose $\sqrt{S}=0.5\,\mathrm{TeV},\,1.0\,\mathrm{TeV},\,5.0\,\mathrm{TeV}$, which, for $\Lambda_{NC}=1.0\,\mathrm{TeV}$, corresponds to $z=0.25,1.0,25.0$. 

\begin{figure}[h]
\hfill\includegraphics[height=2.2in]{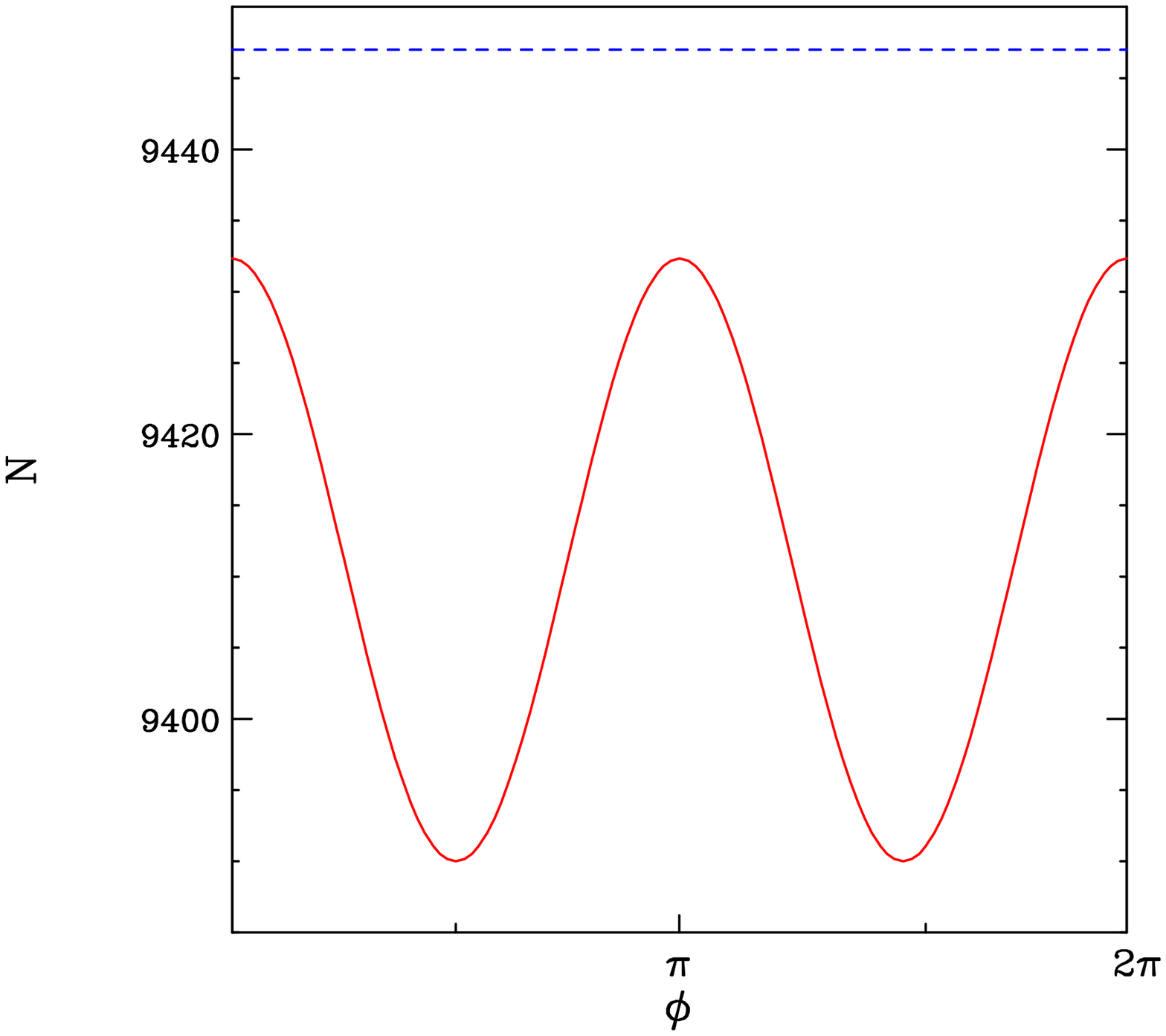}\hfill \includegraphics[height=2.2in]{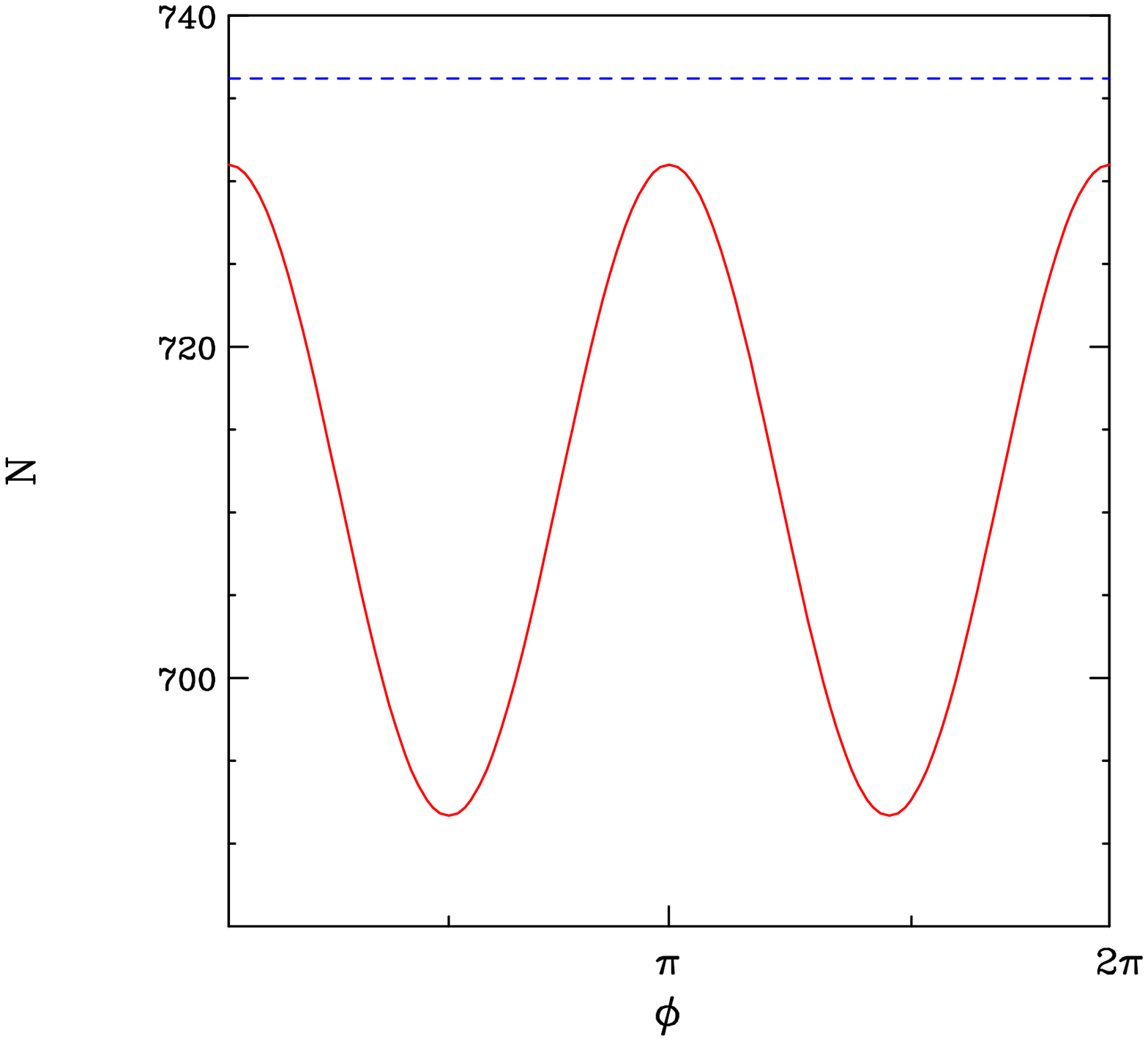}\hspace*{0pt\hfill}
\caption{(color online) In the left panel, the solid line is number of events as a function of $\phi$ for the case $\sqrt{S}=5$\,TeV, $\Lambda_{NC}=1$\,TeV, $\lambda=\pi/4$, $\mathcal{L}=500$\,fb$^{-1}$ and no cut on $\cos\theta$. The dashed line is the uniform background from QED with no $\cos\theta$ cut. The right panel shows the effect of imposing the additional cut $|\cos\theta|\leq 0.9$.\label{dsigdphi}}
\end{figure}

As in the $e^+e^-\to\gamma\gamma$ case, the $\phi$-dependence of the spin averaged cross section is proportional to $z^2$. With no cut imposed on the polar angle $\theta$, we have
\begin{eqnarray}\label{dsdphi}
\frac{d\sigma}{d\phi}&=&\frac{z^2\alpha^3}{\pi S}\left[\frac{2231}{1440}-\frac{\pi^2}{240}-\frac{5}{4}\zeta(3)
-\sin^2\lambda\left(\frac{86141}{69120}-\frac{11\pi^2}{1440}+ \frac{7\log\rho}{1152}-\zeta(3)\right.\right.\nonumber \\
&+&\left.\left.\left(\frac{1963}{1080}+\frac{\pi^2}{360}- \frac{7\log\rho}{1440}-\frac{3}{2}\zeta(3)\right)\cos^2\phi\right)\right]\,.
\end{eqnarray}
The effect of this characteristic $\phi$-dependence is illustrated in the left panel of Fig\,\ref{dsigdphi}.
The signature of $3\gamma$ $\phi$-dependence, a unique feature of NCQED, can be further enhanced by imposing a cut on polar angle $\theta$. This has a rather large effect on the NCQED signal since the QED contribution to the $e^+e^-\to\gamma\gamma\gamma$ cross section is very sharply peaked in the forward a backward directions. The result of the cut $|\cos\theta|\leq 0.9$ is shown in the right panel of Fig\,\ref{dsigdphi}.

We also checked the NCQED corrections to the QED energy and polar angle distributions of one of the photons. While there are some differences in the shapes of the NCQED distributions relative to their QED counterparts, particularly in the energy distribution, using these differences as a test of NCQED appears difficult because of their $z^2$ and energy dependence. The search for $\phi$-dependence remains the best possibility if the $e^+$ and $e^-$ beams are unpolarized.

There are, however, CP violating terms linear in $z$ in the individual helicity amplitudes, as can be seen in Eq.\,(\ref{eliMq}) or in Appendix \ref{Amps}. To probe these terms it is necessary to consider polarization effects.

\subsection{Polarized Cross Sections}

We will confine our discussion to cross sections arising from longitudinally polarized beams. In this case, a typical cross section can be written \cite{m-p}
\begin{eqnarray}
\sigma_{P_{e^-}P_{e^+}}&=&\frac{1}{4}\left[(1+P_{e^-})(1+P_{e^+})\,\sigma_{RR} + (1-P_{e^-})(1-P_{e^+})\,\sigma_{LL}\right. \nonumber \\ 
&& \hspace{4pt}\left.+(1+P_{e^-})(1-P_{e^+})\,\sigma_{RL} +(1-P_{e^-})(1+P_{e^+})\,\sigma_{LR}\right]\,,
\end{eqnarray}
where, for example, $\sigma_{RL}$ denotes the cross section when the $e^-$ beam has pure right-handed polarization ($P_{e^-}=1$) and the $e^+$ beam has pure left-handed polarization ($P_{e^+}=-1$). The remaining cross sections are defined similarly. 

For the process $e^+e^-\to\gamma\gamma\gamma$, amplitudes with $\la=\bar{\la}$ vanish, and we can express the polarized cross section as \cite{m-p}
\begin{eqnarray}
\sigma_{P_{e^-}P_{e^+}}&=&(1-P_{e^-}P_{e^+})\frac{\sigma_{RL}+\sigma_{LR}}{4} \left[1-\frac{P_{e^-}-P_{e^+}}{1-P_{e^-}P_{e^+}} \frac{\sigma_{LR}-\sigma_{RL}}{\sigma_{LR}+\sigma_{RL}}\right]\nonumber \\
&=&(1-P_{e^-}P_{e^+})\sigma_{\rm unpol}\left[1-P_{\rm eff}A_{LR}\right]\,,
\end{eqnarray}
where the effective polarization $P_{\rm eff}$ and the left-right asymmetry $A_{LR}$ are
\begin{eqnarray}
P_{\rm eff} &=&\frac{P_{e^-}-P_{e^+}}{1-P_{e^-}P_{e^+}}\,, \\
A_{LR} &=& \frac{\sigma_{LR}-\sigma_{RL}}{\sigma_{LR}+\sigma_{RL}}\,.
\end{eqnarray}
The left-right asymmetry can be obtained using the squared amplitudes in Appendix \ref{Amps} and Eq.\.(\ref{totXsec}), which results in
\begin{equation}
A_{LR}=-\frac{E_e^2\cos\la}{\Lambda_{NC}^2}\frac{\sb \Big(4\zeta(3) -\ds{\frac{29}{6}}+ \ds{\frac{4\pi^2}{9}}\Big)} {\Big[\sb(\log(\ds{\frac{4E_e^2}{m^2}})-1)^2(\log(\omega)-1)+3\Big]}\,.
\end{equation}
For the process $e^+e^-\to\gamma\gamma\gamma$, the NCQED correction is the main source of a left-right asymmetry. Competing standard model sources of left-right asymmetry such as $Z$ exchange in M\"oller scattering \cite{HPR} are suppressed because they involve loops. Taking $\Lambda_{NC}=1.0$ TeV, Table \ref{alr} shows the cross section values for the cases \cite{HPR} $P_{e^-}=-P_{e^+}=\pm 0.9$ and several values of $\sqrt{s}$ and $\cos\la=1$.
\begin{table}[h]
\begin{tabular}{|c|c|c|}
\hline
\,$\sqrt{s}$ TeV \,&\,$\sigma_{0.9\,-0.9}$ fb\,&\,$\sigma_{-0.9\,0.9}$  fb\,\\
\hline
1.0             & $4315.4$                & $4316.9$                \\
\hline
2.0             & $1187.9$                & $1189.3$                \\
\hline
3.0             & $557.1$                 & $558.6$                 \\
\hline
4.0             & $325.1$                 & $326.6$                 \\
\hline
5.0             & $213.9$                 & $215.4$                 \\
\hline
\end{tabular}
\caption{The cross sections for 90\% left-right and right-left polarized beams are shown for $\Lambda_{NC}=1$\,TeV and $\lambda=0$. \label{alr}}
\end{table}

As the numbers in the Table\,\ref{alr} indicate, the left-right asymmetry, though non-zero, must be distinguished from a fluctuation in the large left-right symmetric QED $e^+e^-\to\gamma\gamma\gamma$ cross section. To obtain a sense of the range of values of $\Lambda_{NC}$ that can be probed polarized cross sections, we examined the signal to square root of background ratio
\begin{equation} \label{bound}
R=\frac{\mathcal{L}(\sigma_{0.9\,-0.9}-\sigma_{-0.9\,0.9})} {\sqrt{\ds \mathcal{L}(\sigma_{0.9\,-0.9}+\sigma_{-0.9\,0.9})}}\,.
\end{equation}
\begin{figure}[h]
\hfill\includegraphics[height=2.2in]{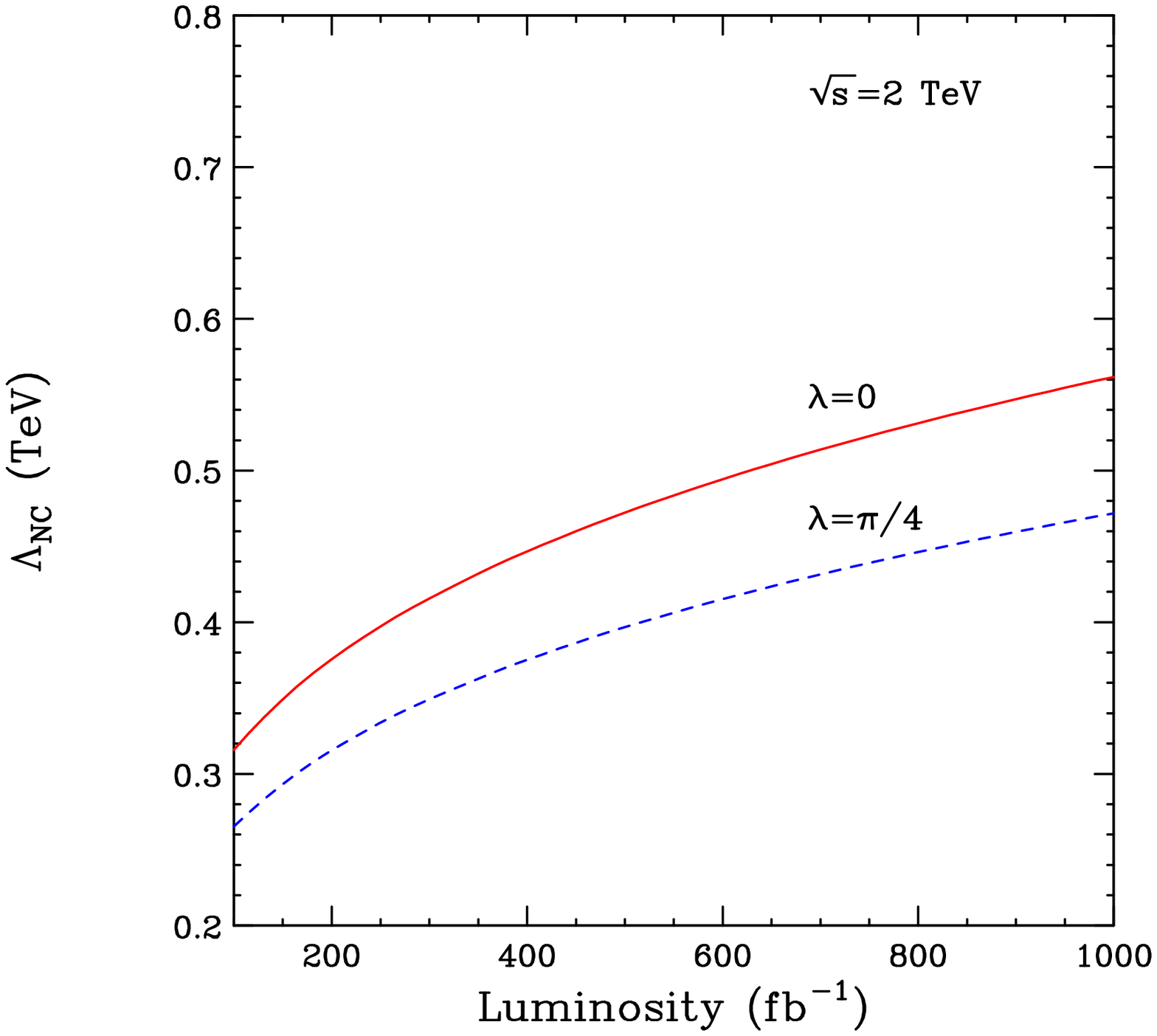}
\hfill\includegraphics[height=2.2in]{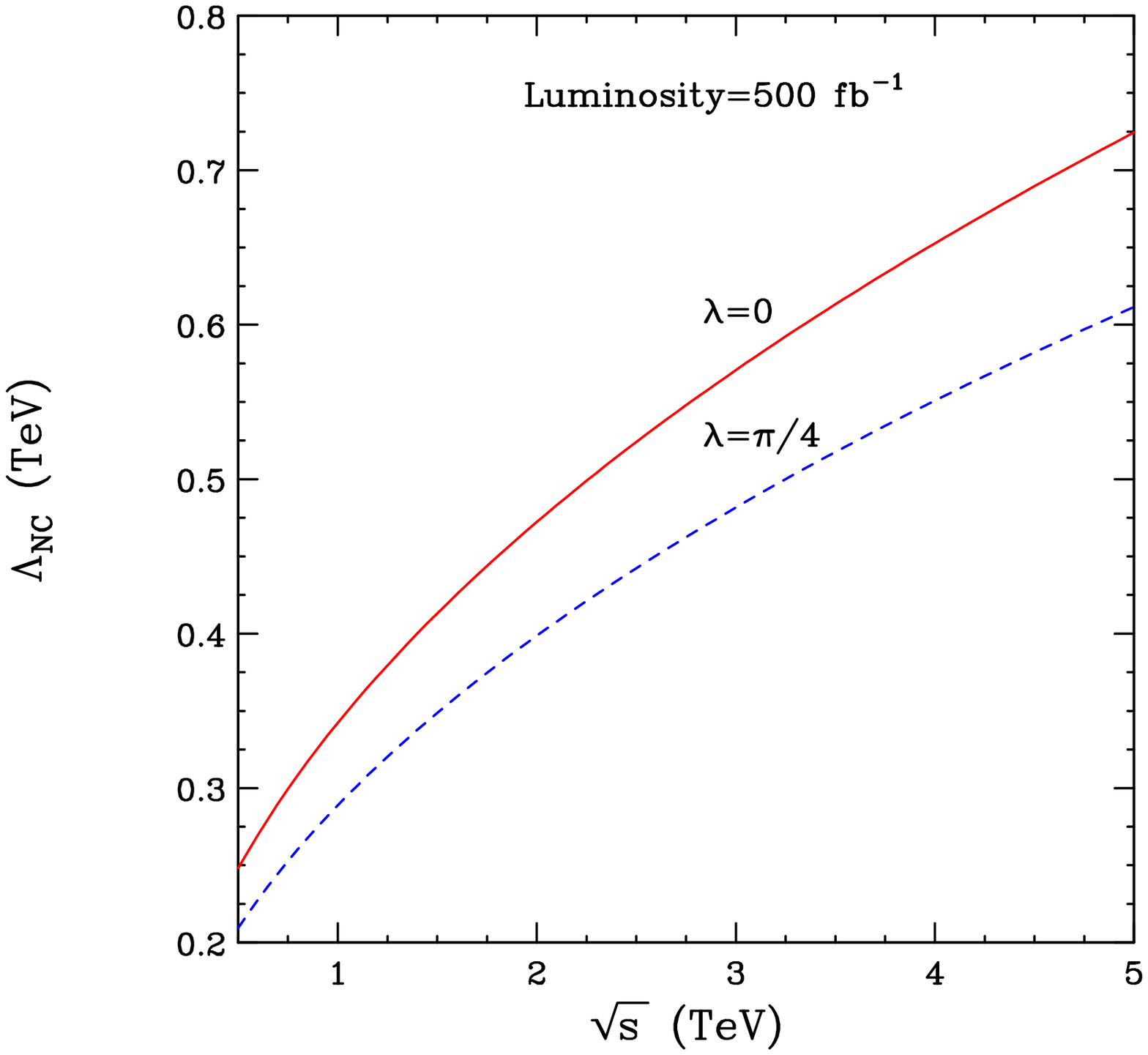}\hspace*{0pt\hfill}
\caption{(color online) The bounds on $\Lambda_{NC}$ attainable using the left-right asymmetry of the total cross section are illustrated as a function of luminosity at $\sqrt{S}=2$ TeV (left) and as a function of $\sqrt{S}$ for $\mathcal{L}=500\,\mathrm{fb}^{-1}$ (right). The solid lines correspond to $\lambda=0$ and the dashed lines to $\lambda=\pi/4$. \label{lamnc}}
\end{figure}
Requiring $R\geq 3$ implies the bounds attainable on $\Lambda_{NC}$ illustrated in Fig.\,\ref{lamnc} as function of luminosity $\mathcal{L}$ and collider energy $\sqrt{S}$.

\begin{figure}[h]
\hfill\includegraphics[height=2.2in]{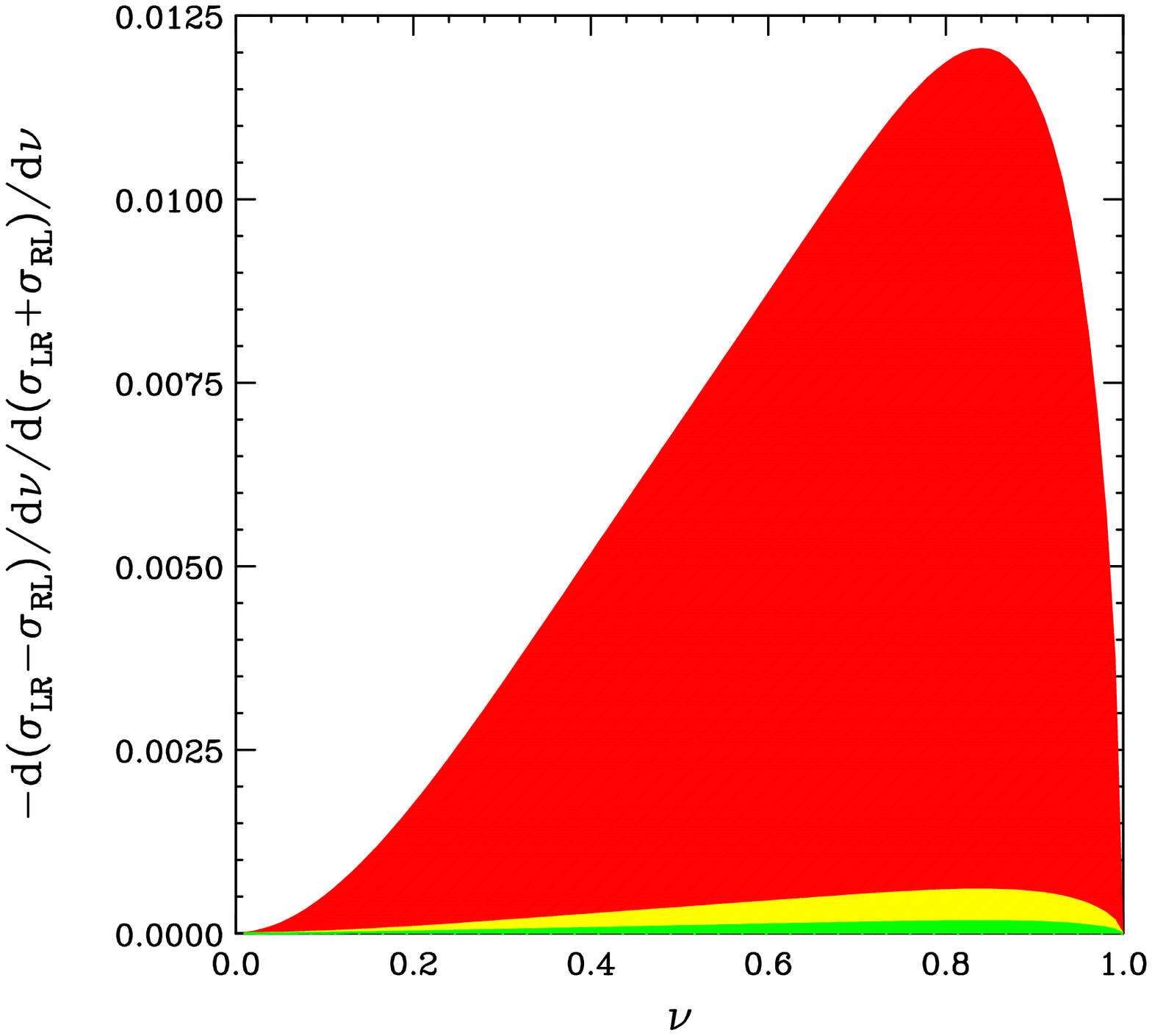}
\hfill\includegraphics[height=2.2in]{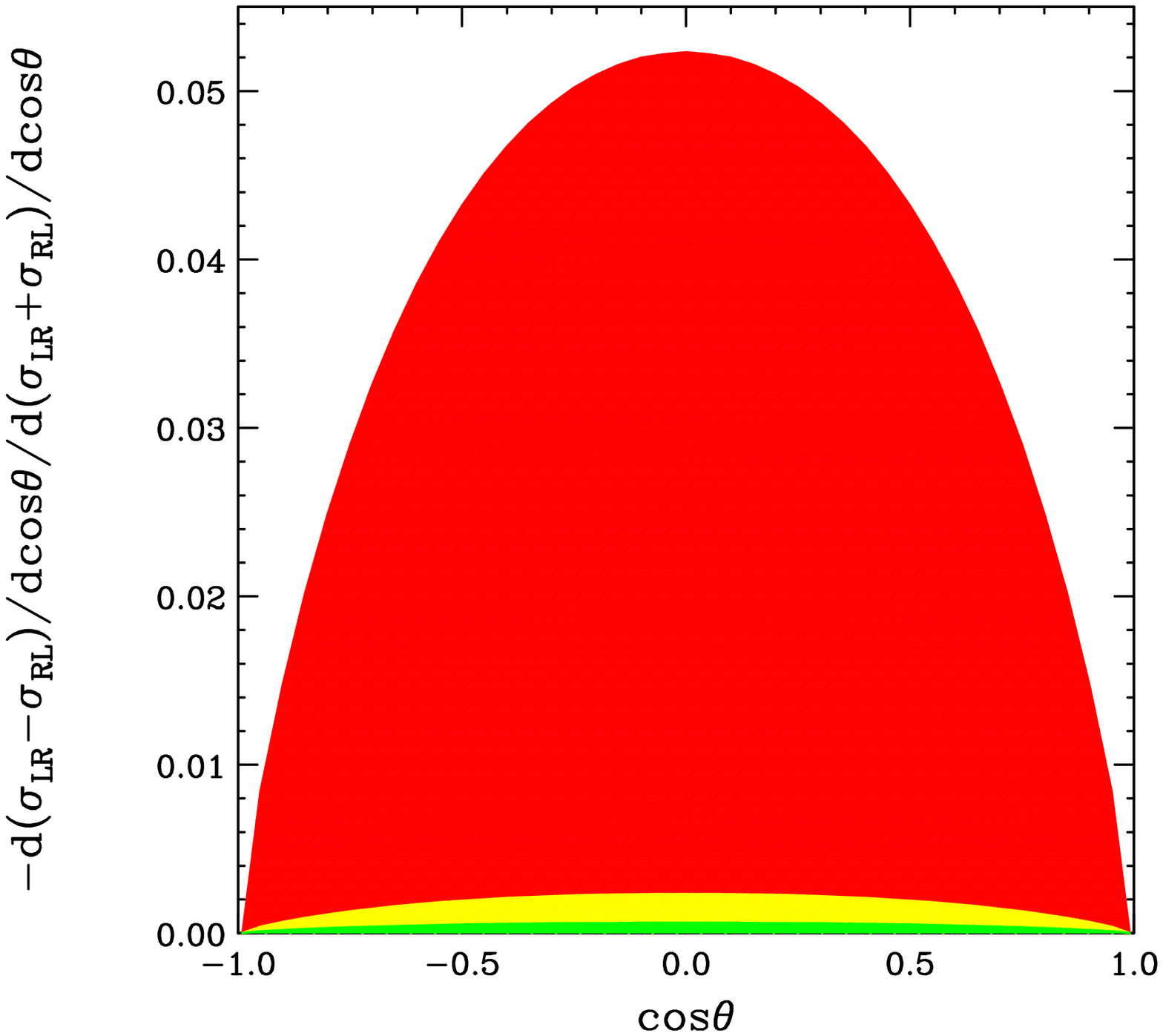}\hspace*{0pt\hfill}
\caption{(color online) The polarization asymmetries with respect to the photon energy fraction $\nu$ (left) and the photon angle with respect to the beam axis (right) are shown. The three shaded regions correspond to center of mass energies of 0.5, 1.0 and 5.0 TeV.  \label{alrez}}
\end{figure}

\begin{figure}[h]
\hfill\includegraphics[height=2.2in]{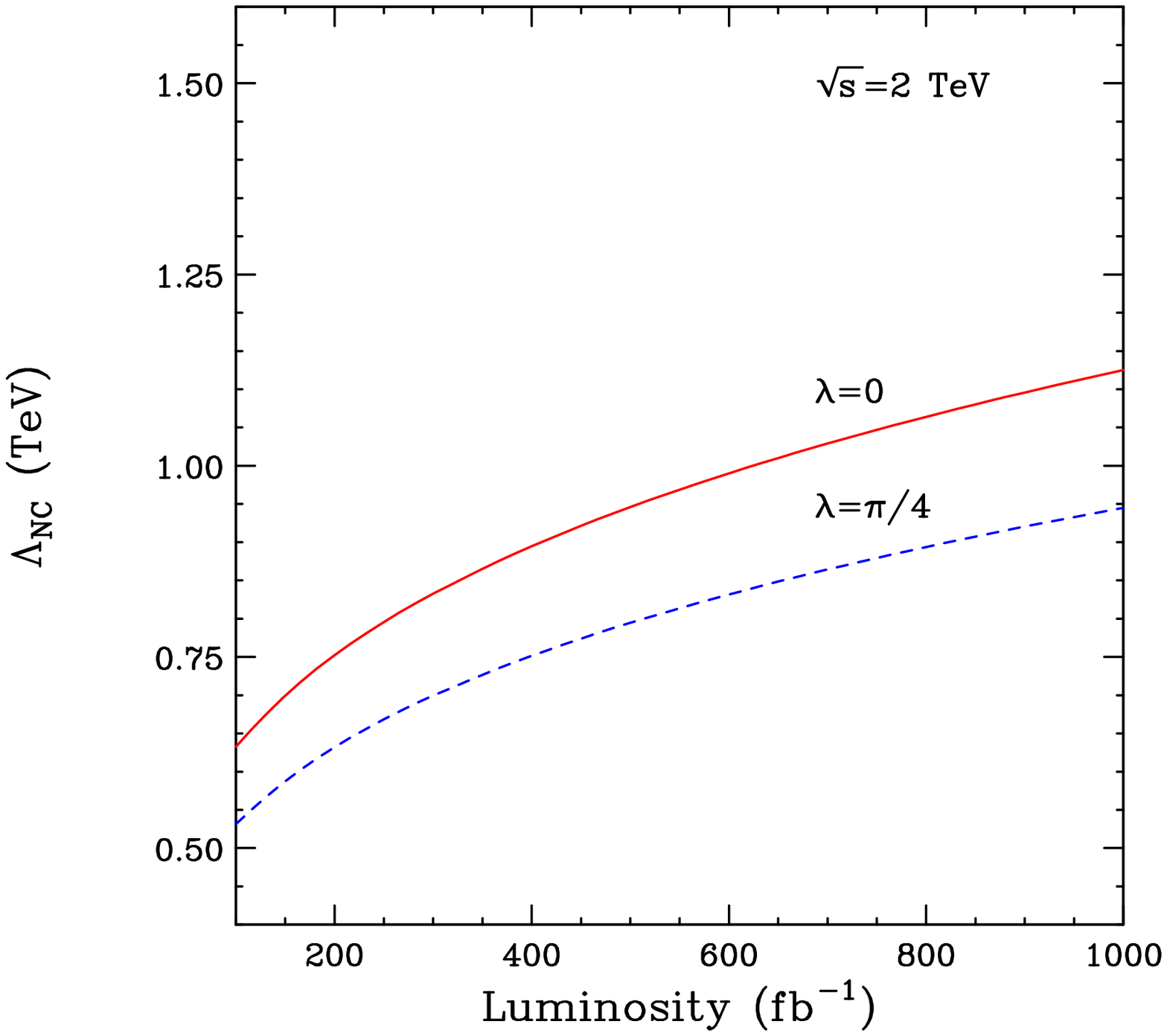}
\hfill\includegraphics[height=2.2in]{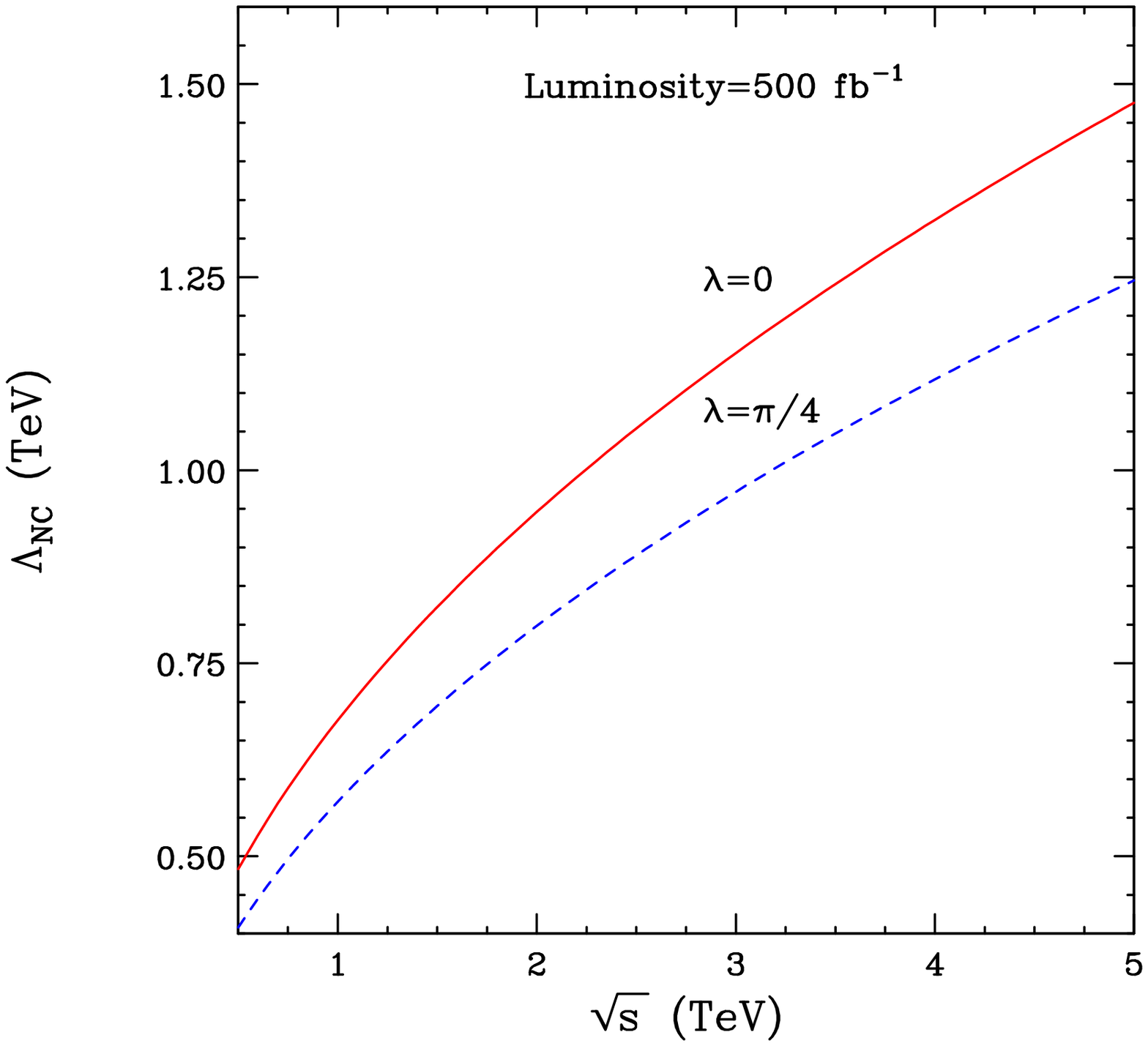}\hspace*{0pt\hfill}
\caption{(color online) Same as Fig.\,\ref{lamnc} with $|\cos\theta|\leq 0.85$. \label{lamnc85}}
\end{figure}

\begin{figure}[h]
\includegraphics[height=2.2in]{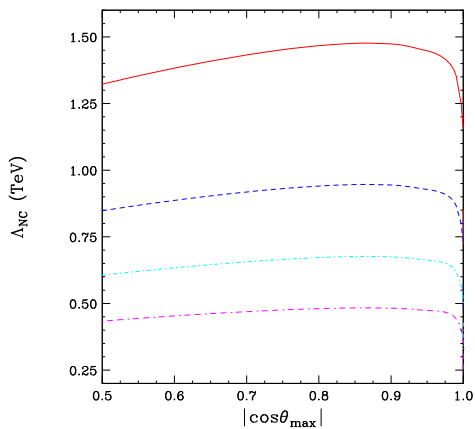}
\caption{(color online) The bounds on $\Lambda_{NC}$ for $0.5\leq|\cos\theta_{\rm max}|\leq 1.0$ are shown for $\sqrt{S}=0.5\,\mathrm{TeV}$ (dot-dash-dash), 1.0 TeV (dot-dash), 2.0 TeV (dashed) and 5.0 TeV (solid). Here $\mathcal{L}=500\,\mathrm{fb}^{-1}$ and $\cos\lambda=1$. The bounds scale as $\mathcal{L}^{1/4}$ and $\sqrt{\cos\lambda}$.\label{lamnc_cuts}}
\end{figure}

The constraints on $\Lambda_{NC}$ obtainable from the polarized total cross section suggest that cuts on the polarization asymmetries in distributions such as
\begin{equation}\label{poldist}
\frac{\sc\ds(d\sigma_{LR}-d\sigma_{RL})/d\nu} {\sc\ds(d\sigma_{LR}+d\sigma_{RL})/d\nu}\quad\mathrm{or}\quad \frac{\sc\ds(d\sigma_{LR}-d\sigma_{RL})/d\cos\theta} {\sc\ds(d\sigma_{LR}+d\sigma_{RL})/d\cos\theta} 
\end{equation}
could improve the bounds on $\Lambda_{NC}$. These distributions are shown in Fig.\,\ref{alrez}.
While both distributions show a distinct left-right asymmetry, the $\cos\theta$ distribution is the most promising from the experimental point of view in that it can be rather large -- $\sim\;\mathrm{few}\;$\% -- over a substantial region of $\cos\theta$. By imposing cuts on $\cos\theta$ it is possible substantially increase the lower bound on $\Lambda_{NC}$ obtained using Eq.\,(\ref{bound}). The largest lower bound is obtained by restricting $\cos\theta$ as $|\cos\theta|\leq 0.85$, which is illustrated in Fig.\,(\ref{lamnc85}).
The behavior of the bound on $\Lambda_{NC}$ as the cut on $\cos\theta$ varies from $|\cos\theta|\leq 0.5$ to $|\cos\theta|\leq 1.0$ is shown in Fig.\,\ref{lamnc_cuts}. 

\section{Discussion and Conclusions}

To summarize, we have computed the noncommutative contributions to the angular and energy distributions of a single photon as well as to the total cross section for the process $e^+e^-\to\gamma\gamma\gamma$ assuming both the unpolarized and polarized electron and positron beams. Because we are dealing with a three particle final state, it is possible to include these corrections using only the space-space portion of the tensor $\theta^{\mu\nu}$. This enables us to avoid the use of the space-time terms $\theta^{k0}$ and thereby satisfy the requirement of unitarity \cite{Chai}. The use of space-time terms cannot be always avoided in $2\to 2$ processes e.g. $e\gamma\to e\gamma$,  $e^+e^-\to\gamma\gamma$ or $e^+e^-\to e^+e^-$ and this tends to complicate their interpretation. The cross sections and distributions depend on the angle $\lambda$ between the beam direction and the non-commutativity vector $\vec{\theta}$. In the unpolarized case, the noncommutative effects are second order in the ratio $z=S/\Lambda_{NC}^2$, whereas in polarized case, the noncommutative effects are leading order in $z$.

In the unpolarized case, the shapes of the QED and NCQED energy distributions are quite different but it is the dependence of the cross section on the azimuthal angle $\phi$ which offers the best opportunity to detect non-commutative effects. The observation of any variation of the cross section with respect to $\phi$ is a clear violation of Lorentz symmetry. It is possible to introduce reasonable cuts significantly enhance this signature of non-commutativity. 

Further, the use of polarized beams makes it possible to probe the order $z$ CP violating terms in the helicity amplitudes by measuring the left-right asymmetry. In contrast to M\"oller scattering, where $Z$ exchange introduces a large standard model left-right asymmetry which competes with the NCQED asymmetry, the NCQED left-right asymmetry in the $3\gamma$ final state is the dominant source of asymmetry, with standard model contributions being suppressed by loops.  Even without cuts on the polarized cross section, the bounds attainable on $\Lambda_{NC}$ are competitive with those obtained in pair annihilation \cite{HPR}. Imposition of cuts on $\cos\theta$, the angle between one of the photons and the beam direction, extends the reach on $\Lambda_{NC}$ to the TeV range. Accumulating enough data to reach these bounds will require monitoring the (unknown) direction of the non-commutativity vector $\vec{\theta}$. Techniques for doing this were proposed by Hewett, Petriello and Rizzo \cite{HPR} and implemented by the OPAL collaboration \cite{Abbiendi}.

Currently, the experimental lower bound on $\Lambda_{NC}$ is $140$ GeV \cite{Abbiendi}, and the calculations of Ref\, \cite{HPR} indicate that $\Lambda_{NC}$ scales of $1.7$ TeV can be probed in M\"oller scattering at a $500$ Gev $e^+e^-$ collider. Like M\"oller scattering, the NCQED contribution to $e^+e^-\to\gamma\gamma\gamma$ can be parameterized solely in terms of the unitarity preserving space-space components of $\theta^{\mu\nu}$. This, together with its NCQED dominant left-right asymmetry signature, makes the three photon process promising candidate in the experimental search for noncommutative effects.

\begin{acknowledgements}
One of us (WWR) wishes to thank INFN, Sezione di Cagliari and Dipartimento di Fisica, Universit\`a di Cagliari for support. This research was supported in part by the National Science Foundation under Grant PHY-0274789 and by M.I.U.R. (Ministero dell'Istruzione, dell'Universit\`a e della Ricerca) under Cofinanziamento PRIN 2003.
\end{acknowledgements}

\appendix
\section{Spin Averaged Cross Section} \label{avgXsec}

The tensor $\theta_{i j}$ can be parameterized in terms of a unit vector $\vec\theta$ and a noncommutativity scale $\Lambda_{NC}$ as
\begin{equation}\label{thvec}
      \theta_{i j}=\frac{1}{\Lambda_{NC}^2} \epsilon_{i j k} \theta^k\,.
\end{equation}
To define the coordinates, we fix the origin at the center of mass, choose the \emph{z} axis parallel to $\vec{p_1}$ and take $\vec\theta$ in the plane \emph{x-z} plane. In this system, $\vec{k_1}$ is defined by its polar angle $\alpha_1$, its azimuthal the angle $\beta_1$ and its energy $E_1$. Similarly $\vec{k_2}$ is defined by its energy $E_2$, its polar angle $\alpha_2$ and, for convenience, an azimuthal angle $\beta_1+\beta_2$.

The phase space integration is given in detail in Appendix \ref{Phase}, where it is shown that, in addition to the variables mentioned above, it is necessary to introduce a minimum photon energy $E_{\rm min}$ to control the infrared singularities. Introducing the dimensionless variables ($i=1,2$)
\begin{eqnarray} \label{nuc}
      \nu_i=\frac{E_i}{E}&&c_i=\cos{\alpha_i} \nonumber \\
      \epsilon=\frac{E_{\rm min}}{E}&&n=\sqrt{1-\frac{m^2}{E^2}}\,,
\end{eqnarray}
the terms in Eq.(\ref{Mq}) can be expressed as
\begin{equation} \label{trasf}
\begin{array}{ccc}
\sb p=\frac{S}{4} \nu_1 \left(1-n c_1\right)& q=\frac{S}{4} \nu_2 \left(1-n c_2\right)& r=\frac{S}{4}\left(2-\nu_1\left(1-n c_1\right)-\nu_2\left(1-n c_2\right)\right)\\
\sb s=\frac{S}{4} \nu_1 \left(1+n c_1\right)& t=\frac{S}{4} \nu_2 \left(1+n c_2\right)& 
u=\frac{S}{4}\left(2-\nu_1\left(1+n c_1\right)-\nu_2\left(1+n c_2\right)\right)
\end{array}\,.
\end{equation}
The total cross section is expressible in terms of these variables as
\begin{eqnarray} \label{cross}
\sigma_{e^+ e^- \rightarrow \gamma\gamma\gamma}&=&\frac{1}{6\,(4 \pi)^5}
\left(\int^{1-\epsilon}_{\epsilon}\mathrm{d} \nu_1\int^{1}_{1-\nu_1}\mathrm{d} \nu_2  
 +\int^1_{1-\epsilon}\mathrm{d} \nu_1\int^{2-\nu_1-\epsilon}_\epsilon \mathrm{d} \nu_2\right)\nonumber \\
 &\times &\int^{1}_{-1}\mathrm{d} c_1\int^{c_+}_{c_-}\mathrm{d} c_2\int^{2 \pi}_{0}\mathrm{d} \beta_1 \ \frac{\left|\mathcal{M}_{e^+ e^- \rightarrow \gamma\gamma\gamma}\right|^2}{\sqrt{\left(c_+ - c_2\right)\left(c_2 - c_-\right)}}
\end{eqnarray}
where 
\begin{equation}
      c_\pm=c_1-2 \frac{c_1 \left(\nu_1+\nu_2-1\right)\mp
\sqrt{\left(1-c_1^2\right)\left(1-\nu_1\right)\left(1-\nu_2\right)\left(\nu_1+\nu_2-1\right)}}{\nu_1 \nu_2}.
\end{equation}

Using the symmetry of Eq.(\ref{Mq}) under a permutation of $\vec{k_1},\vec{k_2},\vec{k_3}$, and neglecting $m^2$ in the numerator of Eq.\,(\ref{Mq}), the commutative contribution to the cross section becomes
\begin{eqnarray} \label{crossC}
\sigma^{QED}_{e^+ e^- \rightarrow \gamma\gamma\gamma}&=&
\frac{2 \alpha^3}{S\,\pi}
\left(\int^{1-\epsilon}_{\epsilon}\mathrm{d} \nu_1\int^{1}_{1-\nu_1}\mathrm{d}\nu_2
+\int^1_{1-\epsilon}\mathrm{d}\nu_1\int^{2-\nu_1-\epsilon}_\epsilon \mathrm{d} \nu_2\right)\int^{1}_{-1}\mathrm{d}c_1\int^{c_+}_{c_-}\mathrm{d}c_2\nonumber \\
 &\times&\frac{\left(2-\nu_1-\nu_2\right)^2+n^2 \left(c_1 \nu_1+c_2 \nu_2\right)^2}{\left(1-n^2 c_1^2\right)\left(1-n^2 c_2^2\right)\sqrt{\left(c_+ - c_2\right)\left(c_2 - c_-\right)}\ \nu_1^2\nu_2^2}\,,
\end{eqnarray}
where the integration over $\beta_1$ has been completed since the process has axial symmetry for $\vec\theta=0$. The integration of Eq.(\ref{crossC}), neglecting terms which vanish for \mbox{$m^2/S\rightarrow0$}, gives \cite{GT} \begin{equation} \label{cQED}
      \sigma^{QED}_{e^+ e^- \rightarrow \gamma\gamma\gamma}=
      \frac{2\alpha^3}{S} \left(\left(\log\rho-1\right)^2\left(\log\omega-1\right)+
      \left(\log\rho-1\right)\log\omega+3-\frac{\pi^2}{3}\right)
\end{equation}
with
\begin{equation}
      \rho=\frac{S}{m^2}\ \ \ \omega=\frac{E^2}{E_{\rm min}^{\,2}}\,.
\end{equation}
As will be seen below, the NCQED correction is small relative to the pure QED cross section. To be certain that the comparison of the two is sensible, we computed the correction to the QED cross section obtained by retaining all the order $m^2$ terms in the numerators of Eq.\,(\ref{Mq}). The calculations are explained in \cite{Berends,Eidelman,Fujimoto}. Our result, which agrees with that of Berends and Kleiss, is 
\begin{equation} \label{mcQED}
      \sigma^{QED}_{e^+ e^- \rightarrow \gamma\gamma\gamma}=
      \frac{2\alpha^3}{S} \left(\left(\log\rho-1\right)^2\left(\log\omega-1\right)+3\right)
\end{equation}

For the noncommutative term, the integrand is no longer invariant with respect to rotations about the \emph{z} axis and it is necessary to consider the $\beta_1$ integration in more detail. From Eq.(\ref{thvec}) we have
\begin{eqnarray}\label{kthk1}
      \varphi&\equiv&{\textstyle\frac{1}{2}}k_1\Dot\theta\Dot k_2
  ={\textstyle\frac{1}{2}}\left(\vec{k_1}\times\vec{k_2}\right)\Dot\vec\theta \nonumber \\
  &=&\frac{E_1E_2}{2\Lambda_{NC}^2}\,\
  \left(\sin{\alpha_1}\left(\sin{\beta_1}\cos{\alpha_2}\sin\lambda
   +\sin{\alpha_2}\sin{\beta_2}\cos\lambda\right)
  -\cos{\alpha_1}\sin{\alpha_2}\sin{\left(\beta_1+\beta_2\right)}\sin\lambda\right)
  \,,
\end{eqnarray}
where $\lambda$ is the angle between $\vec{p_1}$ and $\vec{\theta}$. Assuming $\varphi$ to be small, the noncommutative factor $w^2$ in Eq.\.(\ref{Mq}) becomes  $w^2=\sin^2\!\!\varphi \cong\varphi^2$. Since $\beta_1$ appears only in $w^2$ we can integrate this factor to obtain
\begin{eqnarray} \label{intb}
\int^{2 \pi}_{0}\mathrm{d} \beta_1\,\varphi^2 
&=&\frac{\pi z^2}{64} \left[(1-3 \cos^2\la)\left(4\,c_1 c_2  (\nu_1+\nu_2-1)+\nu_1 \nu_2 (c_1-c_2)^2\right)\right.\nonumber \\ 
& &\left.\hspace{24pt}+8\,\cos^2\la(1-\nu_1)(1-\nu_2)(\nu_1+\nu_2-1)
\right]\,.
\end{eqnarray}

Owing to the additional factors of $E_1$ and $E_2$ in Eq.(\ref{kthk1}), the noncommutative contribution is infrared finite, and it is possible to set $\epsilon=0$ in Eq.\,(\ref{cross}). Then, using Eq.(\ref{intb}) and the symmetry of Eq.(\ref{Mq}) in $k_1,k_2,k_3$, the expression for the noncommutative contribution to the cross section is
\begin{equation} \label{crossNC}
\sigma^{NCQED}_{e^+ e^- \rightarrow \gamma\gamma\gamma}=-\frac{z^2 \alpha^3}{64\,\pi\,S}
\int^{1}_{0}\mathrm{d}\nu_1\int^{1}_{1-\nu_1}\mathrm{d}\nu_2
\int^{1}_{-1}\mathrm{d} c_1\int^{c_+}_{c_-}\mathrm{d} c_2 \frac{\mathcal{N}(c_1,c_2,\nu_1,\nu_2)}{\mathcal{D}(c_1,c_2,\nu_1,\nu_2)}\,,
\end{equation}
where
\begin{eqnarray}\label{NNC} 
\mathcal{N}(c_1,c_2,\nu_1,\nu_2)&=&\left[(2-\nu_1-\nu_2)^2+n^2 (c_1\nu_1+c_2\nu_2)^2\right]
\left[\left(3\left(\nu_1+\nu_2\right)-1\right)\left(2-\nu_1\nu_2\right)-\nu_1^2\nu_2^2\right.\nonumber \\
&+&\left.(2\nu_1\nu_2-7)(\nu_1+\nu_2)^2+4(\nu_1+\nu_2)^3-(\nu_1+\nu_2)^4
  +n^2\left(c_1 c_2\nu_1\nu_2(\nu_1^2+\nu_1\nu_2+\nu_2^2-1)\right.\right.\nonumber \\
 &+&\left.\left.c_1^2\nu_1^2(1-\nu_2+\nu_1^2+\nu_1\nu_2+\nu_2^2-2\nu_1)
  +c_2^2\nu_2^2(1-\nu_1+\nu_1^2+\nu_1\nu_2+\nu_2^2-2\nu_2)\right)\right] \\
&\times&\left[(1-3c_\lambda^2)\nu_1\nu_2\left(4 c_1 c_2 (\nu_1+\nu_2-1)+\nu_1\nu_2(c_1-c_2)^2\right)
 + 8\,c_\lambda^2(1-\nu_1)(1-\nu_2)(\nu_1+\nu_2-1)\right]\,,\nonumber
\end{eqnarray}
and
\begin{equation}\label{DNC}
\mathcal{D}(c_1,c_2,\nu_1,\nu_2)=\nu_1^2\nu_2^2\left(1-\nu_1\right)\left(1-\nu_2\right)\left(\nu_1+\nu_2-1\right)\left(1-n^2 c_1^2\right)\left(1-n^2 c_2^2\right)\sqrt{\left(c_+ -c_2\right)\left(c_2- c_-\right)}\,.
\end{equation}

When Eq.(\ref{crossNC}) is integrated over $\alpha_1$ and $\alpha_2$, we can take $\rho=m^2/S=0$ except for terms involving $\log\rho$, and take $n=1$ in Eq.\,(\ref{nuc}), which does not introduce any collinear divergences. The integration of Eq.(\ref{crossNC}) can be evaluated analytically and the noncommutative contribution is
\begin{equation} \label{epNC}
\sigma^{NcQED}_{e^+ e^- \rightarrow \gamma\gamma\gamma}=\frac{z^2 \alpha^3}{S}\left[\frac{2231}{720}-\frac{\pi^2}{120}-\frac{5}{2}\zeta(3) +\sin^2\la\left(\frac{\pi^2}{80}+\frac{7}{2}\zeta(3)-\frac{148957}{34560}- \frac{7 \log\!\rho}{960}\right)\right]\,.
\end{equation}
From this expression it is clear that production of the three photons depends on the angle between the incident beam and the vector $\vec\theta$, which violates Lorentz invariance. Since the direction of $\vec\theta$ is not known, we average over $\sin^2\la$ to obtain
\begin{equation} \label{NcQED}
\bar{\sigma}^{NCQED}_{e^+ e^- \rightarrow \gamma\gamma\gamma}=\frac{z^2 \alpha^3}{S}\left(\frac{65219}{69120}-\frac{\pi^2}{480}-\frac{3}{4}\zeta(3)
      -\frac{7 \log\!\rho}{1920}\right)\,.
\end{equation}
Adding Eq.(\ref{cQED}) and Eq.(\ref{NcQED}) we get the total cross section
\begin{equation}
\sigma_{e^+ e^- \rightarrow \gamma\gamma\gamma}=\frac{ \alpha^3}{S} \bigg[2\left(\log\rho-1\right)^2\left(\log\omega-1\right)+6 
+z^2\left(\frac{65219}{69120}-\frac{\pi^2}{480}-\frac{3}{4}\zeta(3)
-\frac{7 \log\!\rho}{1920}\right)\bigg]\,.
\end{equation}
\section{Phase Space} \label{Phase}
Taking the center of mass energies of the electron and positron to be $E$ and the flux factor $\left|\vec{v_1}-\vec{v_2}\right|\to2$, the expression for the three body phase space in the electron-positron center of mass is
\begin{eqnarray} \label{phs1}
\mathrm{d}\Gamma &=&\frac{\left(2\pi\right)^4}{3!\,2\left(4 E^2\right)}\delta^4\!\left(p_1+p_2-k_1-k_2-k_3\right)\frac{\mathrm{d}^3 \vec{k_1}}{\left(2\pi\right)^3 2 E_1}\frac{\mathrm{d}^3 \vec{k_2}}{\left(2\pi\right)^3 2 E_2}\frac{\mathrm{d}^3 \vec{k_3}}{\left(2\pi\right)^3 2 E_3}\, \nonumber \\
&=& \frac{1}{3 S\,(4 \pi)^5}
\delta\!\left(2E-E_1-E_2-E_3\right) \delta^3\!\left(\vec{k_1}+\vec{k_2}+\vec{k_3}\right)\frac{\mathrm{d}^3 \vec{k_1}\mathrm{d}^3 \vec{k_2}\mathrm{d}^3 \vec{k_3}}{E_1 E_2 E_3}
\end{eqnarray}
where we have already introduced the symmetry factor $1/3!$.
In spherical coordinates the components of $\vec{k_1}$ and $\vec{k_2}$ can be defined by
\begin{eqnarray} \label{ksph}
\vec{k_1}&=&E_1\left\{\sin{\alpha_1}\cos{\beta_1},\sin{\alpha_1}\sin{\beta_1}, \cos{\alpha_1}\right\}\, \nonumber \\
\vec{k_2}&=&E_2\left\{\sin{\alpha_2}\cos{\left(\beta_1+\beta_2\right)}, \sin{\alpha_2}\sin{\left(\beta_1+\beta_2\right)},\cos{\alpha_2}\right\}\,.
\end{eqnarray}
The integration over $\delta^3\!\left(\vec{k_1}+\vec{k_2}+\vec{k_3}\right)$ gives
\begin{eqnarray} \label{e3}
      E_3&=&\sqrt{\vec{k_3}^2}
      =\sqrt{\left(\vec{k_1}+\vec{k_2}\right)^2}\nonumber \\
      &=&\sqrt{E_1^2+E_2^2+2 E_1 E_2\left(\cos{\alpha_1}\cos{\alpha_2}+\cos{\beta_2}\sin{\alpha_1}\sin{\alpha_2}\right)}
\end{eqnarray}
where we have used the definitions (\ref{ksph}). The phase space integral Eq.(\ref{phs1}) can then be written
\begin{eqnarray} \label{phs2}
\mathrm{d}\Gamma &=& 
\frac{\delta\!\left(2E-E_1-E_2-E_3\right)}{3 S\,(4 \pi)^5}
\frac{\mathrm{d}^3 \vec{k_1}\mathrm{d}^3 \vec{k_2}}{E_1 E_2 E_3} \nonumber \\
&=& \frac{\delta\!\left(2E-E_1-E_2-E_3\right)}{3 S\,(4 \pi)^5}
\mathrm{d}E_1\mathrm{d}E_2\mathrm{d}\cos{\alpha_1}\mathrm{d}\cos{\alpha_2}
\mathrm{d}\beta_1\mathrm{d}\beta_2\frac{E_1 E_2}{E_3}
\end{eqnarray}
with $E_3$ given by Eq.(\ref{e3}). The remaining integration over  $\delta\!\left(2E-E_1-E_2-E_3\right)$ leads to 
\begin{eqnarray} \label{eqdelta}
      E_3^2&=&E_1^2+E_2^2+2 E_1 E_2
\left(\cos{\alpha_1}\cos{\alpha_2}+\cos{\beta_2}\sin{\alpha_1}\sin{\alpha_2}\right)\nonumber \\
 &=&\left(2E-E_1-E_2\right)^2\,.
\end{eqnarray}
Solving Eq.(\ref{eqdelta}) for $\beta_2$ results in two solutions
\begin{equation} \label{beta}
      \beta_{2}=\pm\beta_{20}
\end{equation}
with
\begin{equation} \label{beta20}
      \beta_{20}=\arccos{\left(\frac{E_1 E_2 (1-\cos{\alpha_1} \cos{\alpha_2})-2 E (E_1+E_2-E)}{E_1 E_2 \sin{\alpha_1} \sin{\alpha_2}}\right)}\,.
\end{equation}
The derivative of the argument of the $\delta$ with respect to $\beta_2$,
\begin{equation} \label{derd}
      \frac{d\left(2E-E_1-E_2-E_3\right)}{d\beta_2}=
      \frac{E_1 E_2}{E_3}\sin{\alpha_1}\sin{\alpha_2}\sin{\beta_2}\,.
\end{equation}
then gives 
\begin{equation} \label{delE}
      \delta\!\left(2E-E_1-E_2-E_3\right)= \frac{1}
      {\left|\ds\frac{E_1 E_2}{E_3}\sin{\alpha_1}\sin{\alpha_2}\sin{\beta_{20}}\right|}\left(\sb
      \delta(\beta_2-\beta_{20})+   \delta(\beta_2+\beta_{20})\right)
\end{equation}
Replacing the Eq.(\ref{delE}) in Eq.(\ref{phs2}) we get
\begin{equation} \label{phs3}
d\Gamma=\frac{1}{3 S\,(4 \pi)^5}
\mathrm{d}E_1\mathrm{d}E_2\mathrm{d}\cos{\alpha_1}\mathrm{d}\cos{\alpha_2}
\mathrm{d}\beta_1\mathrm{d}\beta_2\nonumber \\
\frac{\delta\!\left(\beta_2-\beta_{20}\right)+
\delta\!\left(\beta_2+\beta_{20}\right)}{\sin{\alpha_1}\sin{\alpha_2}\sin{\beta_{20}}}\,.
\end{equation}
In the squared amplitude, $\beta_2$ is present only in the  noncommutative factor $w^2$. As  can be seen from Eq.(\ref{trasf}), the other variables are independent of $\beta_2$. The integration over $\beta_2$ produces $w^2(\beta_{20})$ and $w^2(-\beta_{20})$, but these contributions are equal after the subsequent $\beta_1$ integration. Hence, for both the QED and NCQED terms, we can write
\begin{eqnarray}
d\Gamma &=& \frac{2}{3 S\,(4 \pi)^5}
\frac{\mathrm{d}E_1\,\mathrm{d}E_2\,\mathrm{d}\cos{\alpha_1}\,\mathrm{d}\cos{\alpha_2}\, \mathrm{d}\beta_1}{\sin{\alpha_1}\, \sin{\alpha_2}\, \sin{\beta_{2}}}\,,
\end{eqnarray}
where for simplicity we have dropped the zero on $\beta_2$. 
The limits of integration on, say, $\cos\alpha_2$ are constrained by Eq.\,(\ref{beta20}). Using this equation to solve for $\sin\alpha_1\sin\alpha_2\sin\beta_2$, we find, in the notation of Eq.\,(\ref{nuc}),
\begin{equation}
\sin{\alpha_1}\, \sin{\alpha_2}\, \sin{\beta_{2}}= \sqrt{(c_+ - c_2)(c_2 - c_-)}
\end{equation}
with
\begin{equation}
c_\pm=c_1-2 \frac{c_1 \left(\nu_1+\nu_2-1\right)\mp \sqrt{\left(1-c_1^2\right)\left(1-\nu_1\right)\left(1-\nu_2\right) \left(\nu_1+\nu_2-1\right)}}{\nu_1 \nu_2}.
\end{equation}

The limits of integration on the energies are determined by solving the relations
\begin{equation} \label{disE}
      E_{\rm min}\leq E_j\leq E\qquad E_1+E_2+E_3=2 E\,,
\end{equation}
with $j=1,2,3$, for $E_1$ and $E_2$. Elimination of $E_3$ yields the additional inequalities 
\begin{equation} \label{disE2}
      E_2\geq E-E_1\,,\qquad \leq 2E-E_1-E_{\rm min}\,.
\end{equation}
The limits of integration on $E_2$ depend on whether $E_{\rm min}\leq E_1\leq E-E_{\rm min}$ or $E-E_{\rm min}\leq E_1\leq E$. In the former case,
 \begin{equation}
      E-E_1\leq E_2\leq E\,, 
\end{equation}
while in the latter
\begin{equation}
      E_{\rm min}\leq E_2\leq 2 E-E_1-E_{\rm min}\,. 
\end{equation}
\section{Squared Modulus of the Helicity Amplitudes} \label{Amps}

The helicity amplitudes with permuted photon helicities can be derived by the corresponding permutation of the variables $p,q,r$ and $s,t,u$. Amplitudes with every helicity reversed just change the sign of CP-breaking term, while amplitudes in which the electron and positron are exchanged can be derived changing the sign of the antisymmetric term and exchanging the variables $p,q,r$ with $s,t,u$. The amplitudes \mbox{$\{\pm,\pm;\lambda_1,\lambda_2,\lambda_3\}$} and \mbox{$\{\la,\bar{\la};\pm,\pm,\pm\}$} are zero.
The twelve non-zero helicity amplitudes are given in the following table.
\begin{table}[h]
\begin{tabular}{|c|c|c|c|}
\hline
\sb \,$\la,\bar{\la};\la_1,\la_2,\la_3$\, & \,$|\mathcal{M}_{\la,\bar{\la};\la_1,\la_2,\la_3}|^2$\, &  \,$\la,\bar{\la};\la_1,\la_2,\la_3$\, & \,$|\mathcal{M}_{\la,\bar{\la};\la_1,\la_2,\la_3}|^2$\, \\
\hline
\sb $+,-;+,+,-$ & $\ds\frac{r^2}{pqst}\mathcal{A}_-$ & $+,-;-,-,+$ & $\ds\frac{u^2}{pqst}\mathcal{A}_-$ \\
\hline
\sb $+,-;+,-,+$ & $\ds\frac{q^2}{prsu}\mathcal{A}_-$ & $+,-;-,+,-$ & $\ds\frac{t^2}{prsu}\mathcal{A}_-$ \\
\hline
\sb $+,-;-,+,+$ & $\ds\frac{p^2}{qrtu}\mathcal{A}_-$ & $+,-;+,-,-$ & $\ds\frac{s^2}{qrtu}\mathcal{A}_-$ \\
\hline
\sb $-,+;-,-,+$ & $\ds\frac{r^2}{pqst}\mathcal{A}_+$ & $-,+;+,+,-$ & $\ds\frac{u^2}{pqst}\mathcal{A}_+$ \\
\hline
\sb $-,+;-,+,-$ & $\ds\frac{q^2}{prsu}\mathcal{A}_+$ & $-,+;+,-,+$ & $\ds\frac{t^2}{prsu}\mathcal{A}_+$ \\
\hline
\sb $-,+;+,-,-$ & $\ds\frac{p^2}{qrtu}\mathcal{A}_+$ & $-,+;-,+,+$ & $\ds\frac{s^2}{qrtu}\mathcal{A}_+$ \\
\hline
\end{tabular}
\caption{The absolute squares of the twelve non-zero helicity amplitudes, $|\mathcal{M}_{\la,\bar{\la};\la_1,\la_2,\la_3}|^2$, are expressed in terms of the amplitudes $\mathcal{A}_\pm$ given below.}\label{helamps}
\end{table}

The common factor $\mathcal{A}_\pm$ in all the entries in Table\,\ref{helamps} is
\begin{eqnarray}
\mathcal{A}_\pm &=& 2e^6\left[S-4w^2\left(\frac{3}{2}S+\frac{a\,(p^2+s^2)}{bc}+ \frac{b\,(q^2+t^2)}{ac}+\frac{c\,(r^2+u^2)}{ab} \right.\right.\nonumber \\ [4pt]
&&\hspace{6pt}-\left.\left.\frac{pq+(p+s)(q+t)+st}{c}-\frac{pr+(p+s)(r+u)+su}{b}- \frac{qr+(q+t)(r+u)+tu}{a}\right)\right.\nonumber \\
&&\hspace{6pt}\pm\left.4vw\left(\frac{1}{a}+\frac{1}{b}+\frac{1}{c}\right) \ep(k_1,k_2,p_1,p_2))\right]
\end{eqnarray}

\end{document}